# Enhanced Performance of Dye-Sensitized Solar Cells based on TiO$_2$ Nanotube Membranes using Optimized Annealing Profile

F. Mohammadpour,[a] M. Moradi,[a] K. Lee,[b] G. Cha,[b] S. So,[b] A. Kahnt,[c] D. M. Guldi,[c] M. Altomare[b] and P. Schmuki[b,d],†

[a] Department of Physics, Faculty of Science, University of Shiraz, Shiraz 71454, Iran
[b] Department of Material Science and Engineering, WW4-LKO, University of Erlangen-Nuremberg, Martensstrasse 7, D-91058 Erlangen, Germany
[c] University of Erlangen-Nuremberg, Egerlandstrasse 3, D-91058 Erlangen, Germany
[d] Department of Chemistry, King Abdulaziz University, Jeddah, Saudi Arabia
† Corresponding author Email: schmuki@ww.uni-erlangen.de
Tel.: +49-9131-852-7575, Fax: +49-9131-852-7582

**We use free-standing $TiO_2$ nanotube membranes that are transferred onto FTO slides in front-side illuminated dye-sensitized solar cells (DSSCs). We investigate the key parameters for solar cell arrangement of self-ordered anodic $TiO_2$ nanotube layers on the FTO substrate and namely the influence of the annealing procedure on the DSSC light conversion efficiency. The results show that using an optimal temperature annealing profile can significantly enhance the DSSC efficiency (in our case η = 9.8 %), as it leads to a markedly lower density of trapping states in the tube oxide, and thus to strongly improved electron transport properties.**

$TiO_2$-based dye-sensitized solar cells (DSSCs) have attracted much attention in the last decades mainly because of their low cost of fabrication and relatively high efficiency compared to devices based on other inorganic semiconductors. Cell efficiencies up to 12.3 % were reached with Grätzel-type solar cells by using dye-sensitized $TiO_2$ nanoparticle (NP) films as photo-anode.[1,2] Except for an optimal light harvesting by the dye and injection of carriers into the $TiO_2$ scaffold, a key element for the cell efficiency is the transport of electrons through the photo-anode (*vs.* various recombination pathways). In this context, the charge carrier mobility and cell efficiency are inherently limited in a $TiO_2$ nanoparticle (NP) film by the "random walk" electron transport through the network.[3-5] Therefore, one-dimensional (1D) architectures such as high-aspect ratio nanowire,[6,7] nanorod,[8,9] and nanotube[3,10-15] arrays have been investigated as photo-anodes for DSSCs, in expectation of a directional and thus faster electron transport. Anodic $TiO_2$ nanotube (NT) layers have attracted wide attention primarily due to their geometry but also due to their long term stability and facile fabrication. The process of growing tube arrays is based on a simple anodization of a piece of Ti metal under self-organizing electrochemical conditions.[16]

In principle, the resulting aligned nanotubular oxide layers can directly be used in Grätzel type DSSCs adopting a so called "back-side" illumination configuration, where the Ti metal represents the back contact to the dye-sensitized NT layer (see scheme S1(a) in the ESI).

Nevertheless, one would expect a front-side illumination configuration of the $TiO_2$ nanostructures (scheme S1(b) and (c)) to reach higher cell efficiencies compared to a back-side configuration, this because the photo-anode in this case can be irradiated through the FTO glass so that light losses due to light absorption in the electrolyte and in the Pt coating of the counter electrode are minimized.

In order to establish a front-side illumination configuration with anodic $TiO_2$ NT layers, either Ti metal is sputtered on optically transparent conductive substrate glass (FTO) and completely anodized until a transparent NT layer is formed,[17] or NT layers are detached from the Ti substrate as free-standing membranes and are transferred onto FTO slides. Up to now the latter approach, that is, the layer transfer, led to higher cell performance compared to sputtering Ti layers on FTO.[18-20]

A range of different strategies were developed for transferring and arranging tube membranes for their use in DSSCs. Not only tube length and various decoration techniques but also different geometries and configurations, such as open/closed tube bottom and tube bottom up/down (illustrated in scheme S1 in the ESI) were explored.[18,21-25] In all these works, as grown nanotubes are typically converted to anatase by suitable annealing. In spite of this, a thorough study of the annealing conditions of the tubes (including temperature ramping rates) is still missing. This is surprising as earlier reports have shown that the annealing rate can, for instance, drastically affect the tube wall morphology, the chemical properties[26] and the density of grain internal trapping states,[27,28] and thus could be expected to significantly affect the electron transport properties.

Therefore, the present work explores the feasibility to achieve higher efficiencies for front-side illuminated DSSCs, particularly by improving the electron transport properties of the tubes using optimized annealing conditions (specifically addressing also temperature ramping).

In order to evaluate various solar cell configurations, we produced free-standing membranes with a number of transfer techniques, and we optimized the annealing treatment of the nanotube layers. Under optimized thermal treatment that implies, namely, a controlled heating rate, $TiO_2$ NT membranes in an optimal configuration exhibit improved intrinsic electronic properties and thus lead to photo-anodes for front-side illuminated DSSCs with significant increase in their efficiency.

$TiO_2$ nanotube layers were formed by anodization of Ti foils at 60 V for 1 h in an ethylene glycol-based electrolyte (containing 0.15 M of $NH_4F$ and 3 vol.% of DI water). A re-anodization approach (see the ESI for additional details) was used to detach the nanotube layers from the Ti substrate. This allows nanotube membranes to be successfully lifted off and transferred as an intact layer onto $TiO_2$ NP-coated FTO glass.

Screening experiments were carried out by fabricating photo-anodes with 20 µm-thick free-standing tube membranes that were transferred on FTO slides in two different geometries, that are, tube-top-up and tube-top-down configurations, as illustrated in scheme S1(b) and (c). These photo-anodes were crystallized (see XRD data in Fig. S1) and then were used for fabricating DSSCs. The preliminary experiments (see results in Fig. S2) show that the tube-top-down configuration (Fig. 1(a)) of the photo-anode along with annealing temperature of 500 °C represent the experimental conditions leading to cells with higher efficiencies (data are discussed in the ESI).

Therefore, these conditions were adopted for all further experiments. Specifically, the crystallization of the tube membranes at different heating/cooling rates (10, 30 and 60 °C min$^{-1}$) was investigated in detail. In every case, the XRD patterns of different photo-anodes (Fig. S1(a)) show the TiO$_2$ nanotube membranes to consist of pure anatase phase regardless of the heating/cooling rate.

From these layers, the DSSCs were fabricated as reported in the experimental section (see the ESI). A "reference cell" was also fabricated by using a 20 µm-thick TiO$_2$ NP layer as photo-anode instead of the detached nanotube membranes.

Fig. 1(b) shows the J-V curves of the cells and summarizes their photovoltaic characteristics measured under simulated AM 1.5 (100 mW cm$^{-2}$) front-side illumination. Clearly, the cell efficiency is strongly affected by the annealing conditions (ramp rate), that is, a significant enhancement of the photovoltaic performance is observed when increasing the heating rate from 10 to 30 °C min$^{-1}$. In particular, a heating rate of 30 °C min$^{-1}$ leads to a significant improvement of the solar light conversion efficiency up to 9.79 %. All efficiency values reported here were obtained by measuring the cell efficiency under defined active area conditions (see detailed discussion in the ESI) and are in line with IPCE measurements (see Fig. S3 in the ESI).

A further increase of the heating rate up to 60 °C min$^{-1}$ leads to a decrease of the cell performance, although the cell efficiency is still markedly higher than measured for slow ramping conditions (*e.g.*, 10 °C min$^{-1}$). The results in Fig. 1 (b) furthermore show that an optimized annealing of the tube membranes (*e.g.*, heating rate of 30 °C min$^{-1}$) led to cells that were more efficient than a DSSC assembled from a TiO$_2$ NP film (this photo-anode was annealed under the same conditions as adopted for crystallizing the membranes).

A factor that commonly strongly affects the photovoltaic performance is the specific dye-loading, that is, the larger the amount of dye loaded by chemisorption on the TiO$_2$ surface, the higher is the light harvesting efficiency and thus the cell efficiency. However, data in Fig. 1(b) clearly show that the highest dye-loading was measured for the TiO$_2$ NP-based photo-anode (this is in line with commonly observed lower BET areas for tube layers compared to nanoparticles). In other words, the dye-loading cannot be the key to explain the significantly higher efficiency of cells fabricated with rapidly-annealed tubes.

Therefore, IMPS measurements were performed in order to assess the electron transport properties of the different dye-sensitized photo-anodes (Fig. 1(c)). Data in Fig. 1(c) show that significant improvement of the electron transport was obtained for DSSCs that were fabricated with membranes crystallized by rapid annealing (*i.e.*, 30 and 60 °C min$^{-1}$). Similar results were obtained also when performing IMPS measurements on the bare photo-anodes (*i.e.*, without dye-sensitization), this confirming that rapid annealing of the tubes enables improved charge transport properties (see Fig. S4 in the ESI). Noteworthy, the 20 µm-thick TiO$_2$ NP layer used for fabricating the reference cell always showed worse charge transport properties (*i.e.*, slower electron transport and faster charge recombination) compared to the NT-based cells, this regardless of the annealing conditions.

Overall, these results show that the geometric features of the TiO$_2$ photo-anode largely affect the charge carrier mobility which in turn influences the cell efficiency. In other words, the electron transport across a TiO$_2$ photo-anode can largely benefit from the use of one-dimensional TiO$_2$ NT arrays in comparison to slower characteristics observed for NP layers [3-5,10-12]. By taking into consideration that the cell efficiency may strongly relate to the structure of the $TiO_2$ scaffold, a morphological evaluation of the crystalline photo-anodes annealed with different ramping rates was carried out by performing scanning and transmission electron microscopy (SEM and TEM) (see Fig. 2 and additional description in the ESI).

These investigations show that a clear separation of the inner and outer shells of the nanotubes and a corrugated structure of the walls (with clearly visible grain boundaries) are obtained when the tube membranes are annealed with a heating rate of 10 °C min$^{-1}$ (see Fig. 2(a) and (b)). The separation of inner and outer shells could not be observed for rapid annealing and the annealing process merged the two shells instead, so that an apparently "single-walled" tubular structure was formed upon crystallization, as shown in Fig. 2(c)-(f). These results are well in line with previous reports.[26,29,30]

In order to explain the large difference in cell efficiencies and electron transport properties we further investigated the differently annealed tubes by transient absorption spectroscopy based on femtosecond (fs) pump-probe experiments (see the ESI for experimental details).

The results are compiled in Fig. 3(a)-(d). In general, such spectra can be separated into three components (holes, trapped and free electrons – see ESI), this in line with literature on $TiO_2$.[28,31-34] The results of corresponding time constants are given in the table in Fig. 3(e). From the spectra in Fig. 3(a) and (c) it can be clearly seen that the absorption in the NIR region, that is, the absorption related to "free" electrons, is markedly more intense for nanotubes annealed with heating rate of 30 °C min$^{-1}$ than for slowly-annealed tubes. Here, the signal recorded for the former at a time delay of 1 ps is nearly two-fold higher in intensity than that of the latter (black curves in Fig. 3(a) and (c)). In fact, we found that a significantly longer lifetime constant for "free" electrons was obtained for anodic $TiO_2$ nanotubes that underwent crystallization by rapid annealing (see the $\tau_3$ values in Fig. 3(e)), while no relevant difference was observed for hole and trapped-electron lifetimes. Moreover, the transient decay curves in Fig. 3(b) and (d) show that also a slower decay kinetic of the absorption component attributed to the "free" electrons was observed for rapidly-annealed tubes (see the blue-dotted profiles in Fig. 3(b) and (d)).

In other words, the results of fs pump-probe spectroscopy are in full agreement with those obtained by IMPS measurements (*vide supra*), that is, a rapid annealing of anodic $TiO_2$ nanotubes enables superior electronic properties that can be ascribed to a much lower density of trapping states.[27,28] Such enhanced electron transport of rapidly-annealed tubes may be related to the robust, crack-free and "single-walled" morphology. On the contrary, slow annealing led to tubes characterized by relatively large number of cracks and this feature may induce charge recombination and limit the electron lifetime, which in turn detrimentally affect the one-dimensional electron transport.

**Conclusion**

Free-standing $TiO_2$ nanotube membranes were fabricated to be used as photo-anodes in front-side illuminated DSSCs. We showed that several experimental factors strongly affect the solar cell efficiency. In particular, the heating rate used during the crystallization of the photo-anodes largely influenced the structural features of the tubes and their charge carrier transport properties which in turn affect, to a large extent, the solar light conversion efficiency of the device.

**Acknowledgments**

The authors would like to acknowledge ERC, DFG and the Erlangen DFG cluster of excellence for financial support. Helga Hildebrand is gratefully acknowledged for valuable technical help. Also the financial support from the Iran Ministry of Science, Research and Technology is gratefully acknowledged.

**Figures**

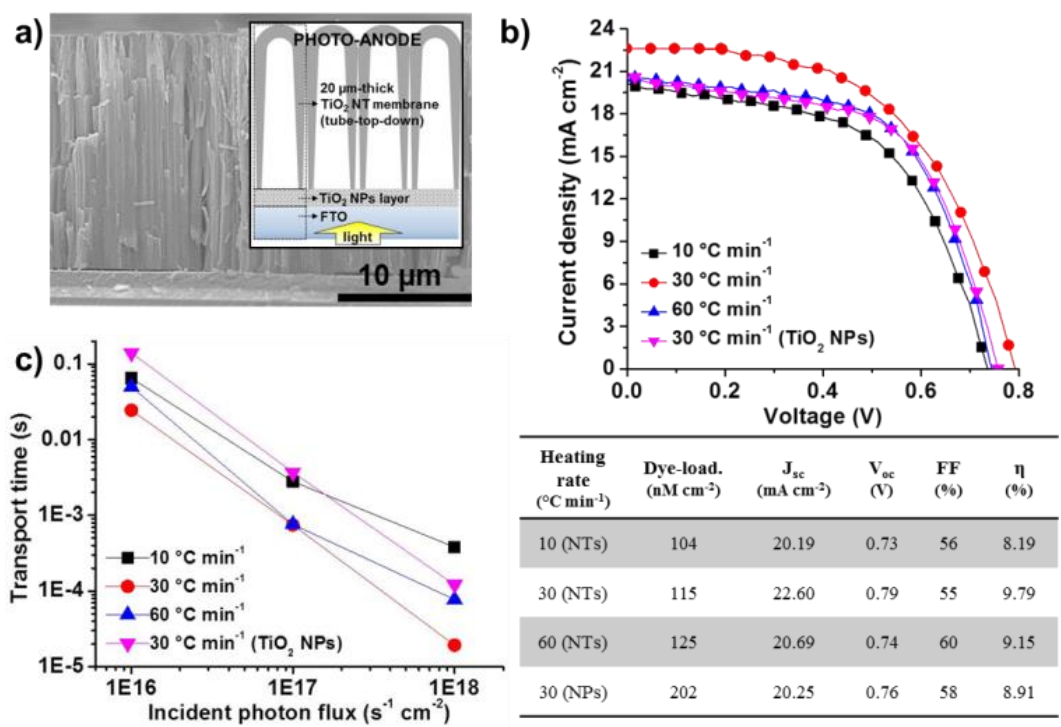

**Figure 1** - a) Cross-sectional SEM image of a photo-anode consisting of a 20 μm-thick $TiO_2$ NT membrane that was fabricated by the re-anodization approach and transferred in tube-top-down configuration on a FTO slide coated with a 2 μm-thick $TiO_2$ NP film. Inset: a sketch of the photo-anode configuration adopted for assembling the DSSCs; b) J-V curves and a summary of the photovoltaic characteristics of the different DSSCs measured under simulated AM 1.5 front-side illumination; c) IMPS measurements performed on the different DSSCs.

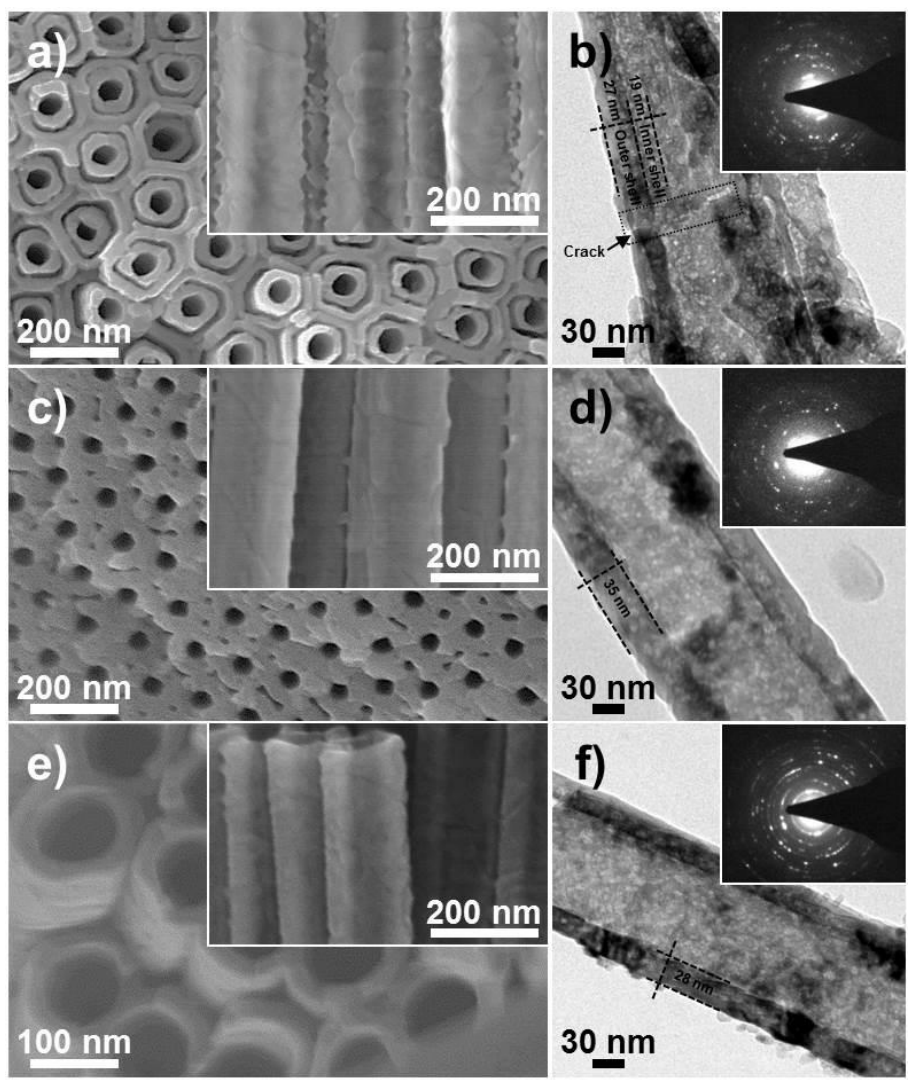

**Figure 2** - High-magnification top view SEM images of TiO$_2$ nanotube membranes annealed in air at 500 °C by using heating rate of a) 10, c) 30 and e) 60 °C min$^{-1}$. The images were taken at middle height along the length of the tubes after cracking the anodic layers. Insets: relative high-magnification cross-sectional SEM images of the nanotubes; TEM images of tubes annealed in air at 500 °C by using heating rate of b) 10, d) 30 and f) 60 °C min$^{-1}$. Insets: relative SAED patterns showing the typical reflections of TiO$_2$ anatase phase.

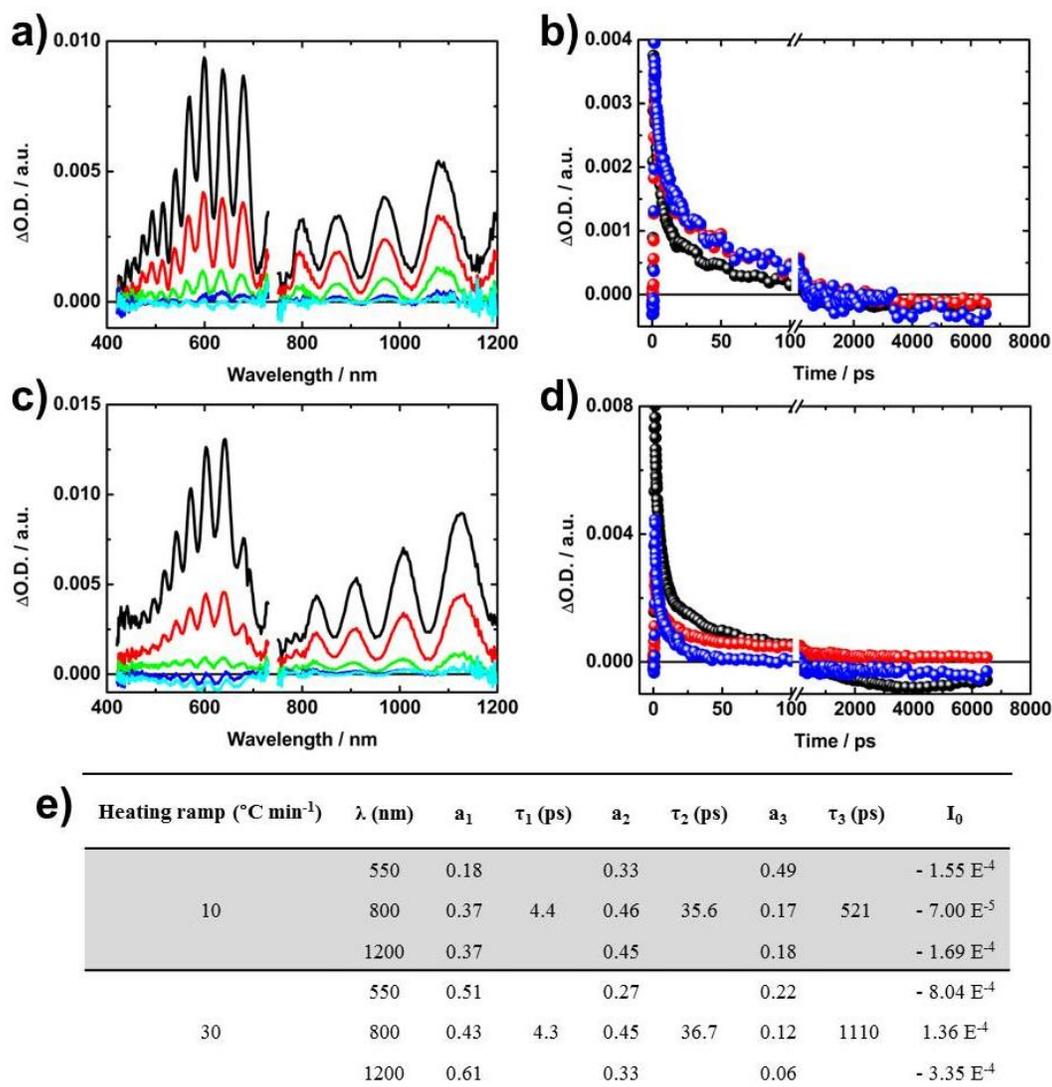

**Figure 3** - a) and c) Differential absorption spectra (visible and near-infrared) measured by femtosecond flash photolysis (258 nm) at different time delays of 1 ps (black), 10 ps (red), 100 ps (green), 1000 ps (blue), and 6500 ps (cyan); b) and d) Relative absorption time profiles of the spectra a) and b) at 550 nm (black), 800 nm (red) and 1200 nm (blue) corresponding to absorption of holes, trapped-electrons and "free" electron, respectively. Data in a), b), c) and d) were measured for *ca.* 1 μm-thick $TiO_2$ nanotube layers (grown on quartz slides) that were annealed at 500 °C with heating rate of a), b) 10 °C min$^{-1}$ and a), b) 30 °C min$^{-1}$; e) data obtained by fitting the decay transient curves to three exponential functions (at 550, 800 and 1250 nm) in order to distinguish the contributions of the different transient species (*i.e.*, holes, trapped-electrons and "free" electron).

Electronic Supplementary Information

**Contents**

- Scheme S1 - Cell configurations
- Anodic growth of $TiO_2$ nanotube layers and membrane fabrication
- Fabrication of photo-anodes and DSSCs
- Characterization of photo-anodes and DSSCs
- Figure S1 - XRD patterns of $TiO_2$ nanotube membrane-based photo-anodes
- Figure S2 - J-V curves and photovoltaic characteristics
- Figure S3 - IPCE spectra of the DSSCs
- Figure S4 - IMPS measurements of bare photo-anodes
- SEM and TEM characterization of $TiO_2$ nanotube membranes
- Figure S5 - XPS analysis
- Figure S6 - EDAX analysis
- Table S1 - Crystallite mean size
- References

**Scheme S1** - Sketch of a) back- and c), d) front-side illuminated TiO$_2$-based dye-sensitized solar cells. In the case of front-side illuminated DSSCs, the TiO$_2$ nanotube membrane can be transferred onto TiO$_2$ NP-coated FTO slides either in b) tube-top-up or c) tube-top-down configuration.

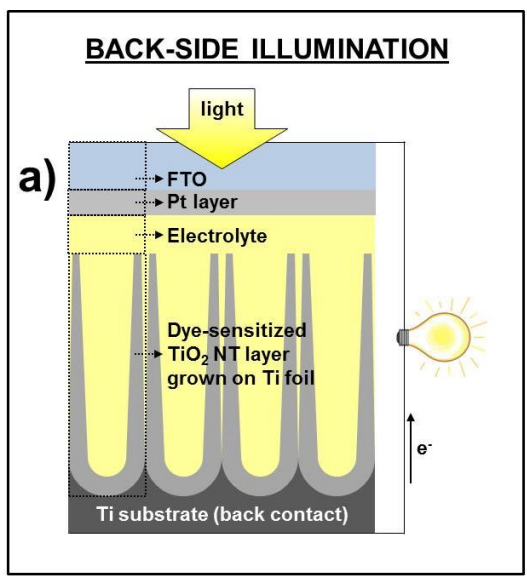

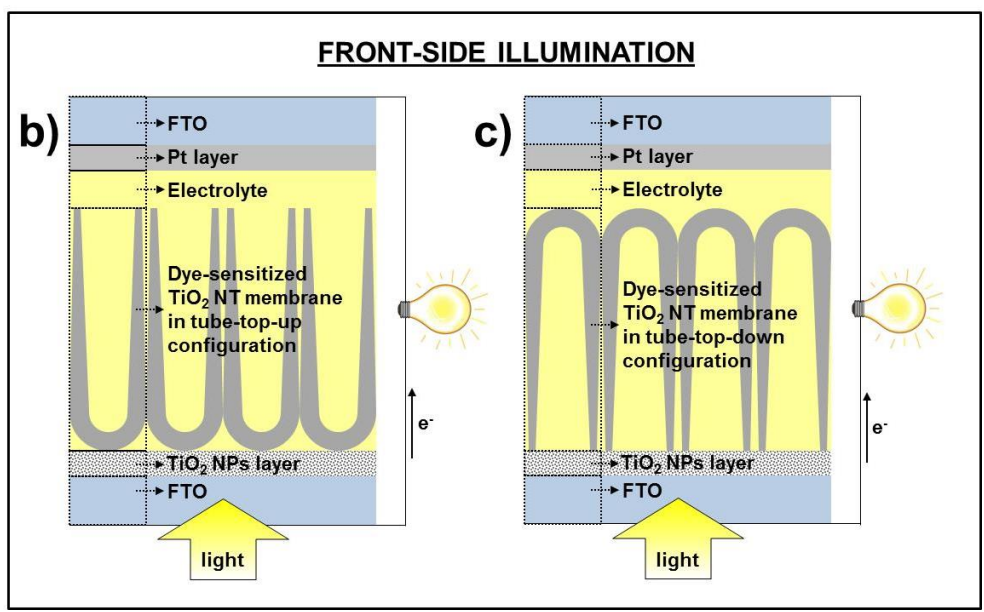

**Anodic growth of TiO$_2$ nanotube layers and membrane fabrication**

Highly ordered TiO$_2$ nanotube arrays were prepared by anodization of Ti foil (0.125 mm thickness, 99.7 % purity, Advent) in a two electrode electrochemical cell by using a Pt sheet as counter electrode. Prior to anodization, the Ti foils were cleaned by sonication in acetone and ethanol (10 min each) followed by rinsing in de-ionized water (DI) and drying in a N$_2$ stream.

The anodization experiments were performed at 60 V at room temperature in an electrolyte composed of ethylene glycol with 0.15 M NH$_4$F and 3 vol.% deionized (DI) water. TiO$_2$ nanotube layers with a thickness of *ca*. 20 µm were produced by anodizing the Ti substrates in the above mentioned conditions for 1 h.

For producing TiO$_2$ nanotube membranes we followed a re-anodization approach introduced by Chen *et al.* [1]. For this, the as-prepared TiO$_2$ nanotube layers were annealed in air at 350 °C for 1 h. Then, the crystallized TiO$_2$ nanotube samples were anodized a second time in the same experimental conditions of the first anodization step. Afterwards, the nanotube layers were detached from the Ti substrate by immersing the tube layers in an aqueous 0.07 M HF solution at 30 °C: the tube layers formed by the second anodization step were still amorphous and therefore underwent preferential chemical dissolution in the HF solution so that the tube layers formed by the first anodization step and pre-annealed could be lifted off as free-standing membranes.

**Fabrication of the photo-anodes and DSSCs**

The photo-anodes were prepared after detachment of the $TiO_2$ nanotube layers by transferring the 20 μm-thick membranes onto FTO slides (7 Ω m$^{-2}$) that were previously coated by doctor-blade technique with a 2 μm-thick film of a commercial $TiO_2$ nanoparticle paste (Ti-Nanoxide HT, Solaronix). The $TiO_2$ nanoparticle layer granted good adhesion, mechanical stability and established good electric contact between tubes and FTO. The membranes were transferred onto FTO either in tube-top-up or tube-top-down (*i.e.*, up-side down) configurations (see scheme S1). A "reference cell" was also fabricated by using a 20 μm-thick $TiO_2$ nanoparticle layer as photo-anode (deposited by doctor-blade technique) instead of the detached nanotube membranes. After drying in air for about 20 min, all the photo-anodes were annealed in air at different temperatures (range of 350-650 °C) for 1 h, with a heating/cooling rate of 10, 30 and 60 °C min$^{-1}$ by using a Rapid Thermal Annealer (Jipelec JetFirst100). Precisely, when screening the different annealing temperatures, a slow heating profile, *i.e.*, 10 °C min$^{-1}$, was always used. On the other hand, when screening the effect of different heating rates (10, 30 and 60 °C min$^{-1}$), all the photo-anodes were annealed at 500 °C.

For fabricating the dye-sensitized cells, the crystalline photo-anodes were firstly immersed into a 300 μM dye solution (D719, Everlight, Taiwan) at 40 °C for 24 h. After dye-sensitization the samples were rinsed with acetonitrile to remove the non-chemisorbed dye and then were dried in a $N_2$ stream. The fabrication of the DSSCs was completed by sandwiching the crystalline dye-sensitized photo-anodes with Pt coated FTO glass as counter electrode using a hot-melt spacer (25μm, Surlyn, Dupont). Finally, the electrolyte (Io-li-tec, ES-0004) was introduced within the interspace of the DSSCs.

**Characterization of the photo-anodes and DSSCs**

For morphological characterization of the TiO$_2$-based photo-anodes, a field-emission scanning electron microscope (FE-SEM, Hitachi SEM FE 4800) and transmission electron microscopy (Philips CM 30 T/STEM microscope) were used.

For determining the crystallographic features of the photo-anodes, X-ray diffraction analysis (XRD) was performed with an X′pert Philips MPD equipped with a Panalytical X´celerator detector using graphite monochromized Cu K$\alpha$ radiation ($\lambda = 1.54056$ Å).

The chemical composition of the TiO$_2$ scaffolds was determined by energy dispersive X-ray analysis (EDAX Genesis) fitted to the Hitachi FE-SEM S4800, and by X-ray photoelectron spectroscopy (XPS, PHI 5600, US).

The current-voltage characteristics of the different DSSCs were measured under simulated AM 1.5 front-side illumination provided by a solar simulator (300 W Xe with optical filter, Solarlight) and by applying an external bias (from - 50 mV up to + 900 mV) to the cell and measuring the generated photocurrent with a Keithley model 2420 digital source meter. Step size and holding time were 23.75 mV and 100 ms, respectively. The active area was defined by the 0.2 cm$^2$-sized opening of the Surlyn seal and a scattering background was used.

Dye loading on the photo-anodes was measured by immersing the different dye-sensitized TiO$_2$ scaffolds in 5 mL of a 10 mM NaOH aqueous solution for 30 minutes. Then the optical absorption of the solutions was measured by a UV-Vis Spectrophotometer (Lambda XLS+, Perkin Elmer).

Intensity modulated photocurrent spectroscopy (IMPS) measurements of the different dye-sensitized TiO$_2$ photo-anodes were carried out in aqueous 0.1 M Na$_2$SO$_4$ solutions by using modulated light (10 % modulation depth) from a high power green LED ($\lambda = 530$ nm). The modulation frequency was controlled by a frequency response analyzer (FRA, Zahner). The light intensity incident on the cell was measured using a calibrated Si photodiode. A three electrode electrochemical cell was used, with a Pt foil as counter electrode and an Ag/AgCl reference electrode.

Incident photon-to-current conversion efficiency (IPCE) measurements were performed with a 150 W Xe arc lamp (LOT-Oriel Instruments) with an Oriel Cornerstone 7400 1/8 m monochromator. The light intensity was measured with an optical power meter.

To study the charge transport process upon photo-excitation of the different TiO$_2$ photo-anodes, transient absorption spectroscopy based on femtosecond (fs) pump-probe experiments were carried out. The 20 μm-thick photo-anodes that were actually used for fabricating the DSSCs could not be employed for these measurements as too thick for transmitting the light pulse, that is, the pulse irradiated on the samples was fully absorbed and no transmitted signal could be measured. Therefore, the pump-probe experiments were performed on *ca*. 3 µm-thick $TiO_2$ nanotube layer grown on quartz slides and crystallized in air at 500 °C with heating rate of 10 or 30 °C min$^{-1}$. Quartz slides as substrates for the tube layers were preferred over FTO glasses since the used excitation wavelength was in the UV range (see below). The tube layers were grown by fully anodizing the 1 µm-thick Ti layers that were evaporated on the quartz substrates. The anodization experiments were performed in the same electrolyte and at the same potential used for fabricating the membranes, and led to $TiO_2$ nanotubes showing comparable features in terms of morphology and phase composition upon crystallization. The pump-probe measurements were performed with a CPA-2110 femtosecond laser (Clark MXR, output 775 nm, 1 kHz, and 150 fs pulse width) using a transient absorption detection system (TAPPS Helios, Ultrafast Systems). The excitation wavelength was generated by third harmonic generation (258 nm) and was characterized by a pulse width of < 150 fs and energy of 200 nJ per pulse.

The obtained transient absorption spectra (reported in Fig. 3(a)-(d) in the main text) are well in line with those typically measured for $TiO_2$ [2-6]. According to the literature, the transient absorption in the visible region (between 400 and 600 nm, this depending on the structural features of the material) is mainly assigned to the holes formed upon photo-excitation. This signal is overlaid by the strong absorption of the trapped electrons that is usually observed in the region between 500 and 900 nm. Furthermore, the transient absorption of the "free" electrons is observed over the entire optical spectrum, although showing particularly strong absorptions in the near infrared (NIR) region. Besides, we also observed the transient absorption/stimulated emission of/from electron-hole recombination induced by the probe light.

The transients that were observed for each sample are represented by the sum of the absorption of the three independent species (holes, trapped and "free" electrons). Therefore, the decay curves of the transients were fitted to three exponential functions (at 550, 800 and 1250 nm). Since all three transient species show overlapping absorptions, global fits were performed in which the three lifetimes are global variables and the non-decaying components were treated as "offset" (a summary of the data is reported in Fig. 3(e) in the main text).

**Figure S1 -** a) XRD patterns of $TiO_2$ nanotube membranes that were transferred onto $TiO_2$ NP-coated FTO slides in tube-top-down configuration and annealed at 500 °C by using different heating rates (10, 30 and 60 °C min-1). XRD patterns of $TiO_2$ nanotube membranes that were transferred onto $TiO_2$ NP-coated FTO slides in b) tube-top-up and c) tube-top-down configurations (see scheme S1(b) and (c)) and annealed in air, at different temperatures (350-650 °C range) for 1 h, with a heating/cooling rate of 10 °C min$^{-1}$.

The XRD patterns measured for these photo-anodes show all the tube membranes to be composed of 100 % anatase phase, this regardless of photo-anode configuration and annealing temperature.

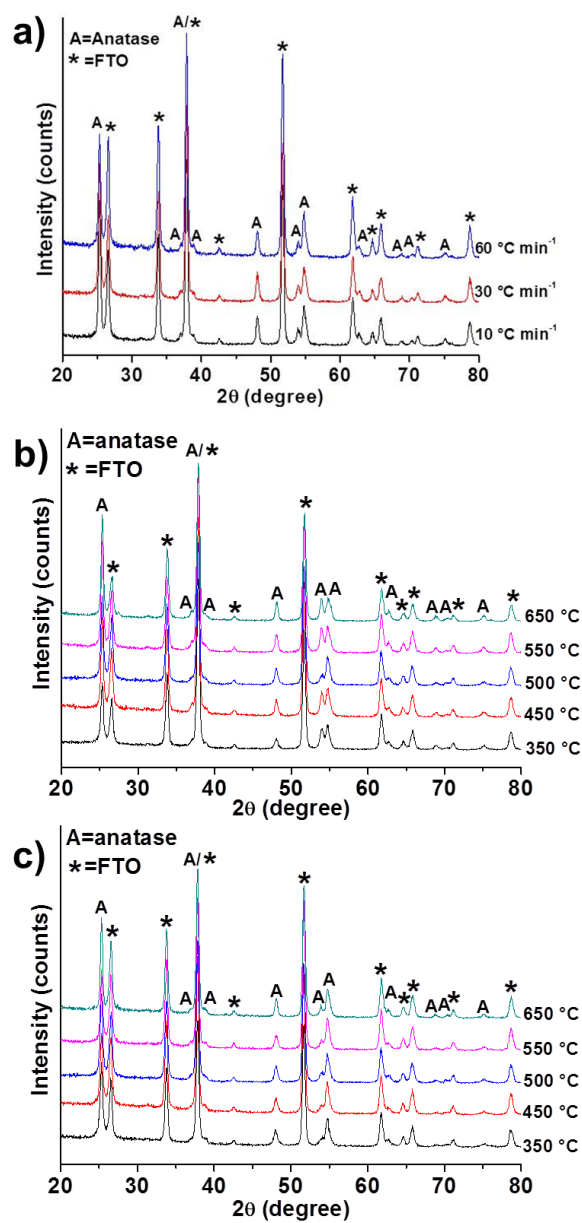

**Figure S2** - a), b) J-V curves and c), d) photovoltaic characteristics of $TiO_2$ nanotube membrane-based DSSCs measured under simulated AM 1.5 illumination. To fabricate the photo-anodes, the $TiO_2$ nanotube membranes were transferred onto $TiO_2$ NP-coated FTO slides in a), c) tube-top-up and b), d) tube-top-down configurations. The photo-anodes were crystallized by annealing in air, at different temperatures (350-650 °C range) for 1 h, with a heating/cooling rate of 10 °C min$^{-1}$. Insets in a) and b) represent a sketch of tube-top-up and tube-top-down configurations, respectively.

These experiments show that the tube-top-down configuration of the photo-anode along with the annealing temperature of 500 °C represent the experimental conditions leading to dye-sensitized solar cells with enhanced efficiencies (*e.g.*, up to 8.19 %). The solar cells mounted with a tube-top-down configuration of the tube membranes lead to significantly higher efficiencies compared to their counterparts (*i.e.*, cell with a tube-top-up configuration) most likely because of the larger dye-loadings obtained with the former geometry.

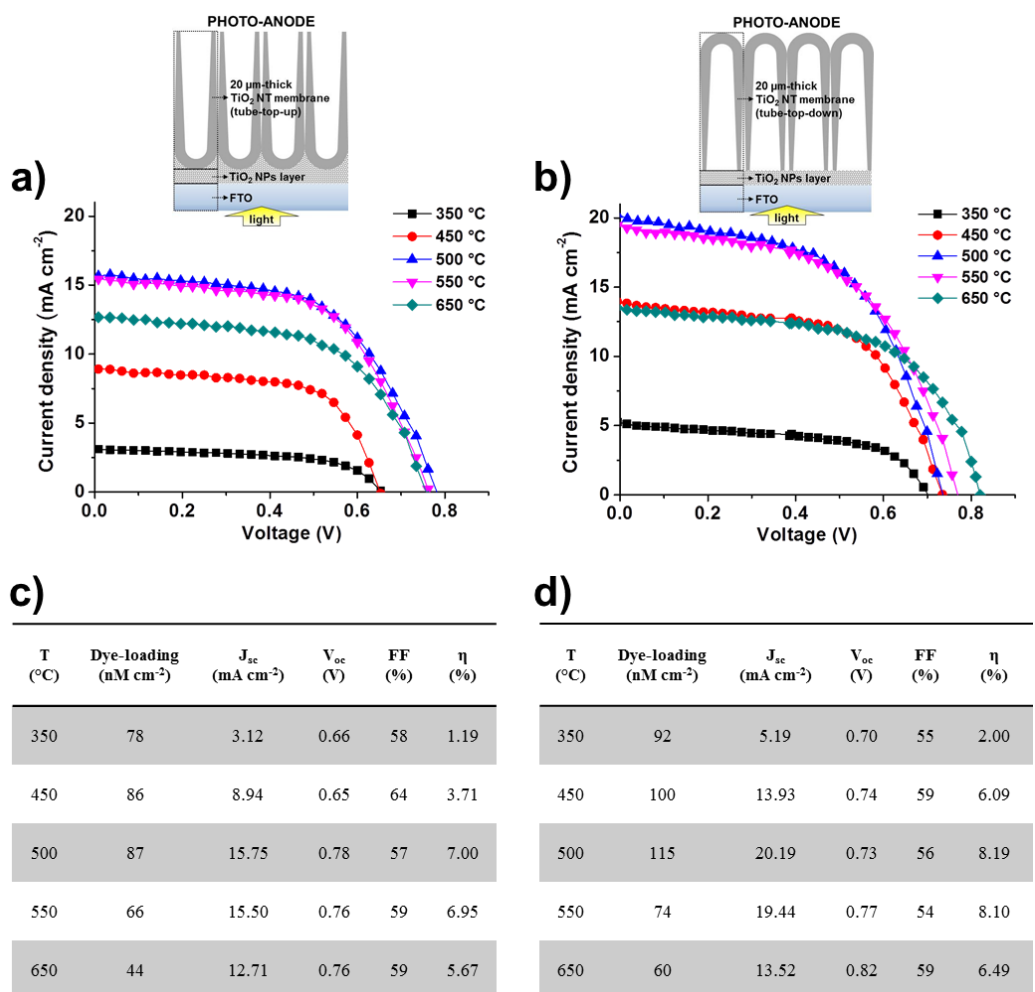

| T (°C) | Dye-loading (nM cm$^{-2}$) | $J_{sc}$ (mA cm$^{-2}$) | $V_{oc}$ (V) | FF (%) | η (%) | T (°C) | Dye-loading (nM cm$^{-2}$) | $J_{sc}$ (mA cm$^{-2}$) | $V_{oc}$ (V) | FF (%) | η (%) |
|---|---|---|---|---|---|---|---|---|---|---|---|
| 350 | 78 | 3.12 | 0.66 | 58 | 1.19 | 350 | 92 | 5.19 | 0.70 | 55 | 2.00 |
| 450 | 86 | 8.94 | 0.65 | 64 | 3.71 | 450 | 100 | 13.93 | 0.74 | 59 | 6.09 |
| 500 | 87 | 15.75 | 0.78 | 57 | 7.00 | 500 | 115 | 20.19 | 0.73 | 56 | 8.19 |
| 550 | 66 | 15.50 | 0.76 | 59 | 6.95 | 550 | 74 | 19.44 | 0.77 | 54 | 8.10 |
| 650 | 44 | 12.71 | 0.76 | 59 | 5.67 | 650 | 60 | 13.52 | 0.82 | 59 | 6.49 |

**Figure S3** - Example of IPCE spectrum (black curve) and relative integrated current density (red curve) of a DSSC the photo-anode of which is fabricated from a $TiO_2$ nanotube membrane transferred onto a $TiO_2$ NP-coated FTO slide (tube-top-down configuration) and annealed at 500 °C (1 h) with a heating/cooling rate of 30 °C min$^{-1}$.

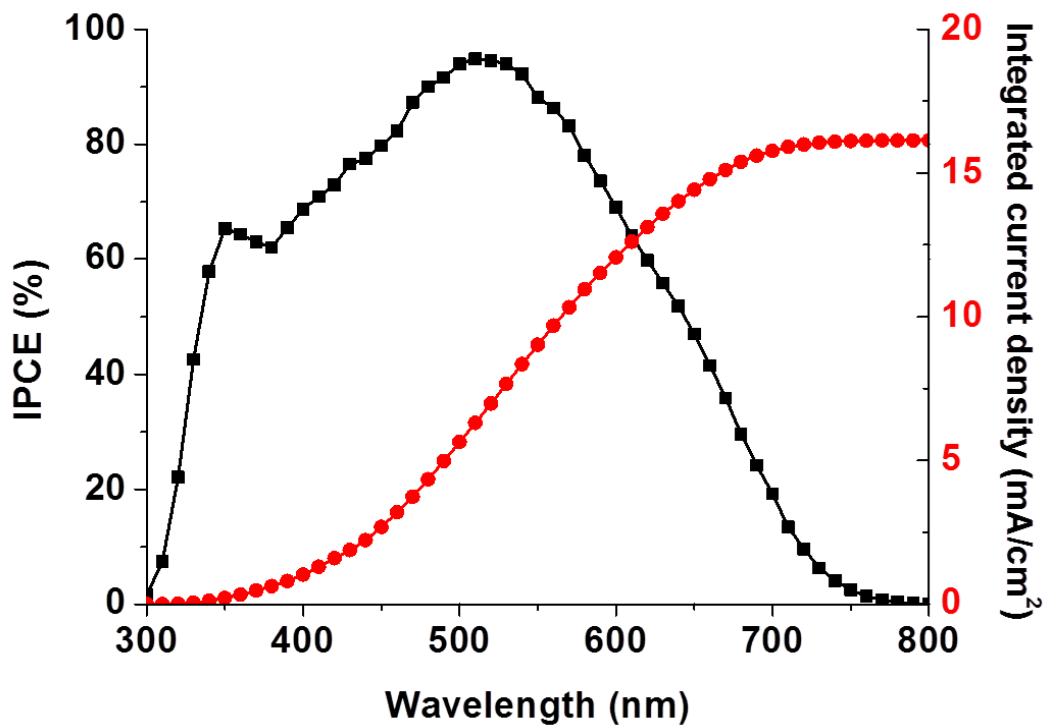

**Figure S4 -** Intensity modulated photocurrent spectroscopy (IMPS) measurements were carried out in aqueous 0.1 M $Na_2SO_4$ solutions on bare (*i.e.*, without dye-sensitization) $TiO_2$ nanotube membrane-based photo-anodes annealed in air at 500 °C with heating rates of 10 and 30 °C min$^{-1}$. A three electrode electrochemical cell was used, with a Pt foil as counter electrode and an Ag/AgCl reference electrode. The experiments were performed by using modulated light (10 % modulation depth) from a high power LED ($\lambda$ = 365 nm). The modulation frequency was controlled by a frequency response analyzer (FRA, Zahner). The light intensity incident on the cell was measured using a calibrated Si photodiode. The photocurrent of the cell was measured using a Zahner electrochemical interface that fed back into the FRA for analysis.

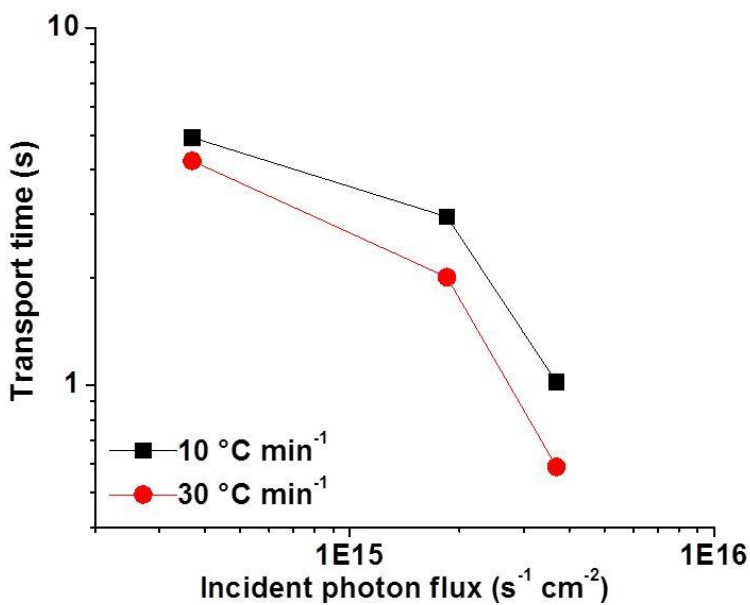

**SEM and TEM characterization of the TiO$_2$ tube membranes**

From the top-view SEM pictures (shown in the main text - Fig 2(a), (c) and (e) - and taken by cracking the anodic film at a middle height along the tube length) it can be seen that all the tubes show an outer diameter of *ca.* 140-160 nm regardless of the annealing conditions. Nevertheless, a heating rate of 10 °C min$^{-1}$ leads to a clear separation of the inner and outer shells of the nanotubes (Fig. 2(a)) [7-9]. This separation could not be observed when tubes were subjected to a more rapid ramping in the annealing process (shown in Fig. 2(c),(e)). In particular, we found that a more rapid annealing led to a merging of the two shells so that an apparently "single-walled" tubular structure was obtained upon crystallization.

Further significant morphological features could be observed by TEM investigations (see Fig 2(b), (d) and (f) in the main text). First, a slow annealing leads to nanotubes with corrugated structure of the walls. This is most likely ascribed to the presence of the inner shell of tube which in this case exhibits a "nanoporous" texture (see Fig. 2(b)) [9]. Second, for these tubes the grain boundaries are clearly visible as fractures along the tube walls (highlighted in Fig. 2(b)). Conversely, the rapidly annealed tubes show a more robust morphology and do not exhibit sharp discontinuities along the tube walls (Fig. 2(d),(f)). Moreover, differences in the tube wall thickness can be seen when comparing tubes crystallized by using different heating rates. The nanotubes crystallized with heating rate of 60 °C min$^{-1}$ were *ca.* 28 nm thick while those annealed at 30 °C min$^{-1}$ were 35 nm thick. In line with this trend, we observed that the tubes annealed by using a rate of 10 °C min$^{-1}$ showed about 46 nm-thick walls, with outer and inner shells of 27 and 19 nm in thickness, respectively.

Besides the effects on the morphology, we also found that the annealing conditions slightly affected both the crystallinity and the chemical composition of the tubes. In fact, as shown by the SAED patterns reported in Fig. 2, the typical reflections of TiO$_2$ anatase phase become brighter (*i.e.*, more intense) when increasing the rate of annealing, meaning that the rapid annealing led to nanotubes with a higher degree of crystallinity. In spite of this, no relevant change of crystallite mean size could be observed for the different annealing conditions (see data in Table S1).

On the other hand, for the chemical composition determined by XPS and EDAX analysis of the different photo-anodes, we found a clearly lower carbon and fluorine contents for rapidly annealed tubes (see Fig. S5 and S6). This may explain the material loss that leads to thinner tube walls, that is, a fast heating rate induces a larger release of carbonaceous species from the inner shell of the tubes (carbon is up-taken in the inner tube shell from the organic-based electrolyte during anodization [8,9]).

**Figure S5 -** XPS analysis of $TiO_2$ nanotube membrane-based photo-anodes annealed in air at 500 °C with heating rates of 10 and 30 °C min$^{-1}$: a) XPS survey; b) high-resolution spectra of the C 1s area; c) summary of the XPS data.

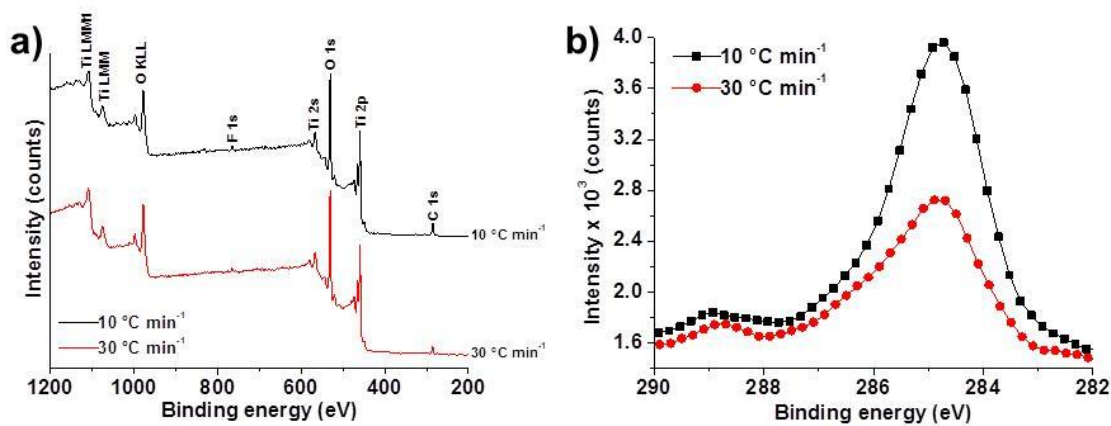

**Figure S6 -** EDAX analysis of $TiO_2$ nanotube membrane-based photo-anodes annealed in air at 500 °C with heating rates of 10 and 30 °C min$^{-1}$: a) and b) EDAX spectra of the photo-anodes annealed with a heating rate of 10 and 30 °C min$^{-1}$, respectively; c) summary of the EDAX data.

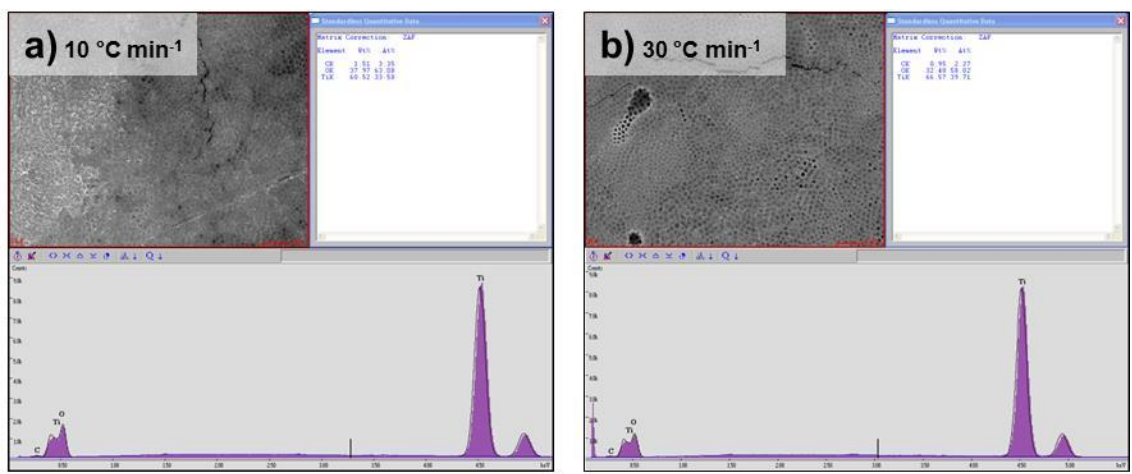

| Heating rate (°C min$^{-1}$) | Element | Wt. % | At. % |
|---|---|---|---|
| | C k | 1.51 | 3.35 |
| 10 | O k | 37.97 | 63.08 |
| | Ti k | 60.52 | 33.58 |
| | C k | 0.95 | 2.27 |
| 30 | O k | 32.48 | 58.02 |
| | Ti k | 66.57 | 39.71 |

**Table S1 -** Mean crystallite size for $TiO_2$ nanotubes annealed in air at 500 °C with heating rates of 10, 30 and 60 °C min$^{-1}$. The crystallite size was determined by: *i)* processing the XRD data by using the Sherrer equation applied to the main reflection of $TiO_2$ anatase phase (2θ = 25.3° corresponding to the (101) crystallographic face); *ii)* TEM investigation (each value represents the average of the size of ten different crystallites).

Average $TiO_2$ grain size for nanotubes annealed by different heating rates

| Heating rate (°C min$^{-1}$) | from XRD data | from TEM analysis |
|---|---|---|
| 10 | 35.6 | 34.0 |
| 30 | 35.9 | 35.0 |
| 60 | 34.2 | - |